\documentclass[runningheads]{llncs}

\usepackage[utf8]{inputenc}

\usepackage{fancyhdr}

\usepackage{graphicx}
\usepackage{float}

\begin{document}
%

\title{Large-scale inference of liver fat with neural networks on UK Biobank body MRI\thanks{This research was supported by a grant from the Swedish Heart- Lung Foundation and the Swedish Research Council (2016-01040, 2019-04756), and used the UK Biobank Resource under application no. 14237.}}

\titlerunning{Large-scale inference of liver fat with neural networks}
%
\author{Taro Langner\inst{1} \and
Robin Strand\inst{1} \and
H\r{a}kan Ahlstr\"{o}m\inst{1,2} \and
Joel Kullberg\inst{1,2}}
\authorrunning{T. Langner et al.}
%
\institute{Uppsala University, 751 85 Uppsala, Sweden \\
\email{taro.langner@surgsci.uu.se} \\    \and
Antaros Medical, BioVenture Hub, 431 53 M\"{o}lndal, Sweden}
\maketitle              

\begin{center}
\textit{Pre-print of accepted submission to MICCAI 2020}
\end{center}

\begin{abstract}

The UK Biobank Imaging Study has acquired medical scans of more than 40,000 volunteer participants. The resulting wealth of anatomical information has been made available for research, together with extensive metadata including measurements of liver fat. These values play an important role in metabolic disease, but are only available for a minority of imaged subjects as their collection requires the careful work of image analysts on dedicated liver MRI. Another UK Biobank protocol is neck-to-knee body MRI for analysis of body composition. The resulting volumes can also quantify fat fractions, even though they were reconstructed with a two- instead of a three-point Dixon technique. In this work, a novel framework for automated inference of liver fat from UK Biobank neck-to-knee body MRI is proposed. A ResNet50 was trained for regression on two-dimensional slices from these scans and the reference values as target, without any need for ground truth segmentations. Once trained, it performs fast, objective, and fully automated predictions that require no manual intervention. On the given data, it closely emulates the reference method, reaching a level of agreement comparable to different gold standard techniques. The network learned to rectify non-linearities in the fat fraction values and identified several outliers in the reference. It outperformed a multi-atlas segmentation baseline and inferred new estimates for all imaged subjects  lacking reference values, expanding the total number of liver fat measurements by factor six.


\end{abstract}
\pagebreak
\section{Introduction}

The UK Biobank has recruited more than 500,000 volunteer participants for medical examination, 100,000 of whom are planned to undergo extensive imaging procedures \cite{sudlow_uk_2015}. Several modalities are involved, including dedicated MRI of the brain, heart, pancreas, and liver, the latter of which enables measurements of accumulated liver fat \cite{wilman2017characterisation}, closely linked to type 2 diabetes and other metabolic disorders. A fat content of $5.5\%$ has been defined as threshold for non-alcoholic fatty liver disease (NAFLD) \cite{browning2004prevalence}, which can progress to non-alcoholic steatohepatitis (NASH) \cite{pagadala2012relevance} and eventually to liver cirrhosis with potentially fatal outcome \cite{younossi2016global}. Both MRI and magnetic resonance spectroscopy (MRS) are non-invasive alternatives to liver biopsy for reliable quantification \cite{reeder2011quantitative}.

Due to its scale, the UK Biobank has the potential to relate liver fat as a biomarker to the wide range of metadata, such as disease outcomes, life-style factors and genetic profiles. At the time of writing, almost 40,000 participants have completed the UK Biobank imaging procedures, but only about 5,000 reference liver fat measurements are available, based on transverse slices acquired from the dedicated liver MRI \cite{wilman2017characterisation}. Like these images, the volumes acquired by neck-to-knee body MRI for body composition analysis \cite{west_feasibility_2016} also include the liver and encode voxel-wise proton density fat fractions (PDFF). Due to using a two- instead of a three-point Dixon technique for reconstruction, these images may encode systematically different PDFF values, and similar protocols have previously shown low agreement with other established methods for quantification of liver fat \cite{kukuk2015comparison}. However, the UK Biobank has released more than 30,000 neck-to-knee body MRI scans, which can be evaluated with machine learning techniques.

Various biological metrics, including liver fat, can be automatically inferred on image data from these scans by convolutional neural networks for regression \cite{Langner2020}. Similar strategies have been previously applied to a range of medical imaging modalities including MRI, with the goal of quantifying properties such as age \cite{cole_predicting_2017-1}\cite{halabi_rsna_2019}\cite{Langner2019} and structures of the heart \cite{xue2017direct}, but also blood pressure and smoking status \cite{poplin2018prediction}. This technique is distinct from neural networks trained for segmentation, which can also be applied for liver fat measurements \cite{irving2017deep}, but typically require ground truth segmentations for training. 

In this work, liver fat measurements are inferred by automated analysis of the more readily available neck-to-knee body MRI with neural networks trained for regression on data from these images \cite{west_feasibility_2016} and the UK Biobank reference values \cite{wilman2017characterisation} as ground truth.
The following contributions are made. 
(1) A specialized framework for liver fat inference from UK Biobank neck-to-knee body MRI, which adapts a generalized system \cite{Langner2020} for superior performance.
(2) A three-way comparison between this method, the reference, and a simple multi-atlas segmentation baseline.
(3) Inferred liver fat values for more than 30,000 UK Biobank subjects, which could be shared for medical research and as a baseline for quality control. 
Code samples and documentation for these implementations are publicly available.\footnote{https://github.com/tarolangner/mri-biometry}

\section{Methodology}

\subsection{UK Biobank Data}

UK Biobank participants were recruited by letter from the National Health Service and scanned at three different imaging centers in the United Kingdom. The majority of subjects reported white British ethnicity ($\sim94\%$) with a mean age of \mbox{64 years} (range 44-82, standard deviation 7.5), mean BMI of 26.6 $kg/m^2$ (standard deviation 4.3) and a share of 52\% males.

\subsubsection{Reference Liver Fat Measurements}

Reference liver fat measurements were available for 4,613 subjects as field 22402-2.0 of the UK Biobank \cite{wilman2017characterisation}. These values are based on the PDFF map of a single transverse slice of the liver, acquired with a Siemens 1.5T MAGNETOM Aera and three-point Dixon technique. The reference method returns the mean PDFF of three manually placed ROIs with liver tissue, avoiding vessels and other confounding structures. To avoid confusion with relative values, liver fat percentage points are referred to as fat fractions (FF). The available values range from 0-46 FF, with a mean of 3.9 $\pm$ 4.6 FF (median 2.1 FF). Of the total 4,613 subjects, 920 (20\%) have recorded liver fat values above $5.5$ FF. 

\subsubsection{Neck-to-knee Body MRI}

The UK Biobank neck-to-knee body MRI scans released as field 20201-2.0 were also acquired with a Siemens 1.5T MAGNETOM Aera device, using a dual-echo Dixon technique with TR=6.69 and TE=2.39/4.77 ms, and flip angle 10deg\cite{west_feasibility_2016}. The resulting water-fat volumes cover most of the body with six overlapping stations, excluding the head and lower legs, whereas the arms and other tissues at the edges of the magnetic field are typically subject to strong image artifacts. Likewise, the borders between stations often contain motion artifacts that can affect the shape of the liver.

\subsection{Experimental Setup}

A neural network was trained for regression of liver fat values on the neck-to-knee body MRI of those subjects with available reference values. It was first evaluated in 10-fold cross-validation and then, after training on the full dataset, applied to the remaining subjects for inference. In both phases, independent measurements from a simple multi-atlas segmentation strategy described in one of the following sections served as a baseline.

\subsubsection{Datasets}

All 32,323 neck-to-knee body MRI scans that have been released at the time of writing were quality controlled by an operator who visually inspected two-dimensional mean intensity projections of the water and fat signals \cite{Langner2020}. Due to water-fat swaps, noise, metal objects, unusual positioning and other artifacts, about 3,6\% of the subjects were excluded, leaving 31,171 images for the experiments. 

Three datasets were formed from those subjects that passed the quality control. Validation dataset \textit{A} consists of those 4,418 subjects with existing reference values as ground truth for the training and validation of the network in 10-fold cross-validation. The atlas technique was also validated on this set. 
Inference dataset \textit{B} consists of the remaining 26,753 quality-controlled subjects for whom no reference values were available. The network was applied to these subjects, but without reference measurements (and consequently no ground truth values) this dataset could not be used as a true independent test set. Instead, the atlas was applied to extract baseline measurements for comparison, but was too slow to process them all within the given time. 
Therefore, comparison dataset \textit{C} was formed as a random subset of 1,000 subjects from dataset \textit{B} and used to compare the network and atlas.

\subsubsection{Image Formatting}

The stations of the neck-to-knee body MRI were fused and resampled to a common spatial resolution of \mbox{2.23mm $\times$ 2.23mm $\times$ 3mm}, with \mbox{370 $\times$ 224 $\times$ 174 voxels}. Next, water and fat fraction images were calculated by voxel-wise division of the water or fat signal intensity by the sum of both signals. This sum image was also used to generate body masks by applying a threshold, calculated as the mean of Otsu filter thresholds for all coronal slices of the summed signal for a given subject.

\begin{figure}
	\centering
	\includegraphics[width=\textwidth]{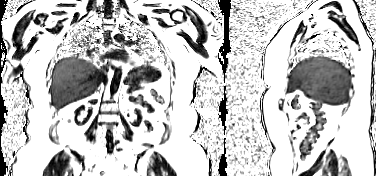}	
	\caption{Input format for the network as extracted from the neck-to-knee body MRI volumes. The upper halves of specific coronal and sagittal fat fraction slices are combined in a two-dimensional 8bit image of $376 \times 176$ pixels, encoding fractions of 0-50\%. For the shown subject, the reference method lists a liver fat percentage of 15\%.}	
	\label{fig_input}
\end{figure}

For the neural network, a highly compressed two-dimensional format with coronal and sagittal fat fraction slices was extracted, as seen in Fig. \ref{fig_input}. Based on the body mask, the coronal slice was extracted at center of mass and the sagittal slice at a quarter of mass, typically locating the latter along the center of the right thigh. Both slices were cropped to exclude the bed and the bottom half of the body and then concatenated. The resulting image of \mbox{376 $\times$ 176} pixels was then compressed to an 8bit format, with fat fractions ranging from 0 to 50\%. No body mask was applied to this format. While there is no guarantee that this strategy captures actual liver tissue it operated robustly and required only about 5 seconds per subject with a GPU implementation.

\subsubsection{Neural Network Configuration}

The convolutional neural network is based on a ResNet50 \cite{he_deep_2016} modified for regression of a single value. Most hyperparameters of the generalized regression framework were retained \cite{Langner2020}, with batch size 32, online-augmentation by translations and weights pretrained on ImageNet. Accordingly, a mean squared error loss was optimized with Adam and a base learning rate of 0.0001 was reduced by factor ten in the last 1,000 iterations of 6,000 iterations in total. Each training sample consists of the two-dimensional image format as extracted from the neck-to-knee body MRI as input and the reference liver fat measurement of the same subject as ground truth value. Training on one split required about 25 minutes whereas prediction was almost instantaneous. All experiments ran on a Nvidia GTX 1080 Ti 11GB GPU, a 12-core Intel Xeon W-2133, 3.60 GHz CPU and 32GB RAM in PyTorch.

\subsubsection{Multi-Atlas Baseline}

A simple multi-atlas segmentation pipeline was implemented on the neck-to-knee body MRI, performing a median readout of minimal liver segmentations.
Three subjects (one female, two male) with high observed variance in liver shape were manually segmented to serve as templates, using the water fraction and signal images to outline the liver and exclude vessels and adjacent tissue.
When applying the atlas, these templates were transformed to each target with a graph-cut based deformable registration technique \cite{ekstrom_fast_2018} on the full paired water and fat fraction volumes, after applying the body mask to remove background noise. A resolution pyramid of six levels was used together with a GPU-implementation of the normalized cross-correlation. The obtained deformation field for each template was applied to the corresponding liver segmentation. The thus aligned binary segmentations were then multiplied and subsequently eroded with a spherical kernel of seven voxels in diameter. The median fat fraction value of the remaining selected voxels was then returned. 
PDFF values of the neck-to-knee body MRI may not directly correspond to those reconstructed by the reference method \cite{kukuk2015comparison}. As a final step, the raw atlas output was therefore fit to the reference values on the validation dataset \textit{A} by linear regression. When applied to other datasets, the same parameters were applied. 


\subsubsection{Evaluation}

The network was trained to emulate the reference method by regression. The success of this training was therefore evaluated by quantifying the quality of fit with the coefficient of determination R$^2$ and reporting the mean absolute error (MAE) and the 95\% limits of agreement (LoA). Furthermore, Pearson's coefficient of correlation $r$ is used to examine the randomness of errors by network and atlas relative to the reference.
The measured values can also be thresholded to identify subjects above the NAFLD risk level of $5.5$ FF. The area under curve (AUC) of the receiver operating characteristic (ROC) curve is reported together with sensitivity and specificity for thresholding at this level.
For some outliers, manual segmentations similar to the atlas templates were created. Their extent was more conservative, and their median FF values are reported after correction with the same linear regression parameters as applied for the atlas.

\section{Results and Discussion}

\begin{figure}
	\includegraphics[width=\textwidth]{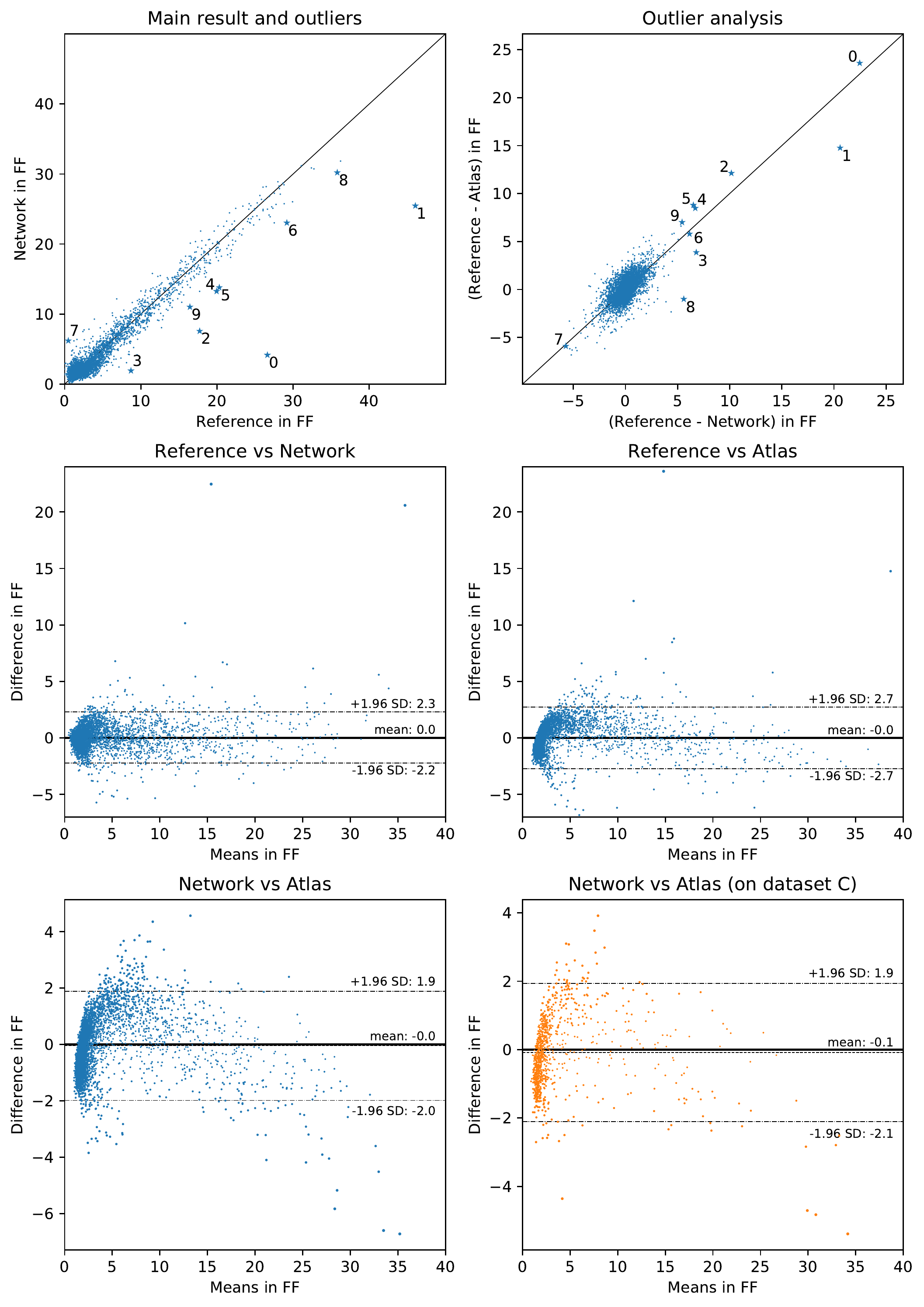}	
	\caption{Results on datasets \textit{A} (blue, N=4,418) and \textit{C} (red, N=1,000). Differences were calculated by subtracting the second named method from the first. Dashed lines denote 95\% limits of agreement. Both network and atlas independently contradict the reference in the annotated top ten network prediction outliers, indicating implausible reference values. Network inference on dataset \textit{C} retains its agreement with the atlas as observed on dataset \textit{A}, showing no signs of deteriorated performance in generalization.
}
	\label{fig_results}	
\end{figure}

Results are shown in Fig. \ref{fig_results} and Table \ref{tab_results}. The network outperformed the atlas on validation dataset \textit{A} but retained the same pattern of agreement with it on comparison dataset \textit{C}, indicating robust generalization to those subjects lacking reference values. On dataset \textit{A}, the network inferred outliers of up to 23 FF, with 16 subjects reaching errors above 5 FF. However, manual segmentation of the top ten outliers yielded measurements that agree for LoA of \mbox{(-0.7 to 1.2 FF)} to the atlas and \mbox{(-5.4 to 7.3 FF)} with the network, but only \mbox{(-23.5 to 8.4 FF)} with the reference. This substantial disparity shows a mismatch between the reference and liver fat as observed in the neck-to-knee body MRI in these subjects that can not be explained by the different imaging protocol alone, but might instead be the result of spurious outliers in the reference method. The errors of atlas and network relative to the reference on dataset \textit{A} are highly correlated ($r=0.715$ with \mbox{LoA $-1.9$ to $2.0$}).

\begin{table*}[b]
	\begin{center}
		\label{tab_results}	
		\caption{Method comparison on datasets \textit{A} (first three rows) and \textit{C} (bottom row). The three final columns assume thresholding at $5.5$ FF, with the first named method as ground truth. MAE: mean absolute error, LoA: 95\% limits of agreement, ROC-AUC: area under receiver operating characteristic curve, Sens: sensitivity, Spec: specificity.}		
		\setlength{\tabcolsep}{3.5pt}
		\begin{tabular}{lcl|ccc|ccc}			
			\hline
			& &  & MAE & R$^2$ & LoA & ROC-AUC & Sens & Spec \\
			\hline
			Reference  & vs & Network & 0.77 & 0.940 & (-2.22 to 2.31) & 0.992 & 89.3 & 98.2 \\
			Reference  & vs & Atlas & 1.03 & 0.912 & (-2.73 to 2.73) & 0.991 & 78.0 & 99.2 \\
			Atlas  & vs & Network & 0.76 & 0.952 & (-1.89 to 1.98) & 0.995 & 97.4 & 95.8 \\
			\hline 
			Atlas (C)  & vs & Network (C) & 0.80 & 0.950 &  (-1.94 to 2.11) & 0.991 & 93.6 & 95.6 \\
			\hline
		\end{tabular}		
	\end{center}
\end{table*}

The agreement between network and reference with LoA \mbox{(-2.2 to 2.3 FF)} is well within the LoA of \mbox{(-4.0 to 3.4 FF)} reported between the gold standard modalities of MRI and MRS \cite{yokoo2018linearity}.
Even within MRI, variability with LoA of \mbox{(-2.2 to 1.8 FF)} for comparable protocols and \mbox{(-2.5 to 1.6)} between different imaging sites has to be expected \cite{hernando2017multisite}.
The atlas was about fifty times slower than the network and affected by systematic differences between the two- and three-point Dixon techniques \cite{kukuk2015comparison}. It was only evaluated after correction with the parameters from linear regression \mbox{($0.9x - 0.8$ FF)}, which failed to resolve a non-linear structure in the range of 1-4 FF, however. Perhaps due to the outliers, the corrected values still tend to overestimate low and underestimate high values. In contrast, the network learned to rectify values in the lower ranges and emulated the reference method better, possibly with information from additional image features. With LoA of \mbox{(-2.7 to 2.7 FF)}, the atlas still surpasses the the generalized regression framework, which used mean intensity projections of water and fat signals for LoA of \mbox{(-4.0 to 4.2 FF)} on the same UK Biobank subjects \cite{Langner2020}.


Several limitations apply to the presented results. The comparison on dataset \textit{C} indicates robust generalization to UK Biobank subjects from any of the given centers with the chosen protocol. However, without an independent test set no conclusions can be drawn for other demographic groups, scanning devices or imaging protocols, which would likely at least require retraining of the network. 
The slice selection strategy is also not guaranteed to be optimal or even certain to capture any liver tissue, and inherently limited by encoding only values from 0-50 FF in 8bit.
Conceptually, the potential of deep learning on the neck-to-knee body MRI is still not fully leveraged, as design choices regarding the compression of the volumes into two-dimensional representations still strongly affects the results, reminiscent of hand-crafted features.
The proposed regression technique also generates no output segmentations, so that individual predictions can not be easily explained or corrected. In contrast, neural networks for segmentation have been proposed for automation of parts of the reference method \cite{irving2017deep} and could provide these, reducing any manual work to the initial creation of representative ground truth segmentations for training only.

However, apart from the underlying design decisions, the presented approach eliminates the need for manual intervention entirely. No human guidance, model-based assumptions or representative ground truth segmentations are required, and inference for thousands of subjects can be performed within a day. Future work will consist in expanding the training and evaluation samples, once available, which is likely to further improve performance \cite{Langner2019}. Likewise, the planned repeat scans of up to 70,000 subjects could potentially be processed without any further changes to the presented system, enabling convenient longitudinal samples of liver fat. The reference method also provides a liver inflammation factor and iron content, and while there is no indication that these could be inferred from the neck-to-knee body MRI in the same way, the presented approach could also be evaluated on the dedicated liver MRI slices directly. Furthermore, the inferred values of dataset \textit{B} could be examined more thoroughly, using known correlations to the metadata and by expanding the baseline measurements.

It is worth noting that after the completion of these experiments an alternative reference set of about 10,000 image-based liver fat measurements has been released by the UK Biobank as field 22436-2.0. A full analysis of these values is beyond the scope of this work, but preliminary results indicate that field 22436-2.0 may be of higher quality than field 22402-2.0 used here. Like the network and atlas, the new reference contradicts field 22402-2.0 on the top ten outliers found in this work. Early attempts to emulate this new reference with the proposed method yielded a closer fit. As the data collection is continuously progressing, future work will be able to examine the relationship between both references and the proposed technique in ever increasing sample sizes.

\section{Conclusion}

In conclusion, the proposed framework can emulate the reference measurements and infer similar liver fat values from the neck-to-knee body MRI of the UK Biobank. It outperforms the atlas and combines high speed with accurate and objective predictions, while eliminating any need for manual intervention or guidance and leaving no room to subjective variability. The inferred liver fat measurements are readily available at large scale for distribution as return data by the UK Biobank. They could be used to identify potentially erroneous outliers as observed in the reference method, but also as an approximation in medical research, enabling larger sample sizes until all subjects have been evaluated with more established gold standard techniques.

\bibliographystyle{splncs04}
\bibliography{references}

\end{document}